\documentclass[aps,twocolumn,epsfig,graphics,showpacs,floatfix,mathbbm,nopacs,hyperref]{revtex4}

\usepackage{amsmath,amsfonts,amssymb,amsthm,stmaryrd,graphics,graphicx,epsfig,times,bbm,psfrag}
\usepackage[usenames]{color}

\begin{document}

\bibliographystyle{apsrev}

\newcommand{\tr}{\operatorname{tr}}
\newcommand{\uinvnorm}{|\kern-2pt|\kern-2pt|}
\newcommand{\wt}{\operatorname{wt}}
\newcommand{\spectrum}{\operatorname{sp}}
\newcommand{\erf}{\operatorname{erf}}
\newcommand{\erfc}{\operatorname{erfc}}
\newcommand{\supp}{\operatorname{supp}}
\newcommand{\diam}{\operatorname{diam}}

\bibliographystyle{apsrev}

\newcommand{\me}{\mathrm{e}}
\newcommand{\mi}{\mathrm{i}}
\newcommand{\md}{\mathrm{d}}
\renewcommand{\vec}[1]{\text{\boldmath$#1$}}

\newcommand{\cc}{\mathbb{C}}
\newcommand{\nn}{\mathbb{N}}
\newcommand{\rr}{\mathbb{R}}
\newcommand{\zz}{\mathbb{Z}}
\newcommand{\id}{\mathbb{I}}

\newtheorem{definition}{Definition}
\newtheorem{theorem}{Theorem}
\newtheorem{lemma}{Lemma}
\newtheorem{corollary}{Corollary}
\newtheorem{property}{Property}
\newtheorem{proposition}{Proposition}
\newtheorem{remark}{Remark}
\newtheorem{example}{Example}
\newtheorem{assumption}{Assumption}

\setlength{\parskip}{2pt}

\newcommand{\identity}{\openone}
\newcommand{\be}{\begin{equation*}}
\newcommand{\bea}{\begin{eqnarray}}
\newcommand{\eea}{\end{eqnarray}}
\newcommand{\ee}{\end{equation*}}
\newcommand{\bra}[1]{\mbox{$\langle #1 |$}}
\newcommand{\ket}[1]{\mbox{$| #1 \rangle$}}
\newcommand{\braket}[2]{\mbox{$\langle #1  | #2 \rangle$}}
\newcommand{\proj}[1]{\mbox{$|#1\rangle \!\langle #1 |$}}
\newcommand{\ev}[1]{\mbox{$\langle #1 \rangle$}}

\newcommand{\suc}[1]{{\color{Red} #1}}

\def\sign{\mbox{sgn}}
\def\H{{\cal H}}
\def\C{{\cal C}}
\def\E{{\cal E}}
\def\O{{\cal O}}
\def\B{{\cal B}}
\def\one{\ensuremath{\hbox{$\mathrm I$\kern-.6em$\mathrm 1$}}}
\def\tr{ \mbox{tr}}

\bibliographystyle{unsrt}

\title{Do mixtures of bosonic and fermionic
atoms adiabatically heat up in optical lattices?}

\author{M.\ Cramer,$^{1}$ S.\ Ospelkaus,$^{2,}$\footnote{Present address: JILA, University of Colorado, Boulder, CO 80309, USA} C.\ Ospelkaus,$^{2,}$\footnote{Present address: NIST Boulder, 325 Broadway, Boulder, CO 80305, USA} K.\ Bongs,$^{2}$ K.\ Sengstock,$^{2}$ and J.\ Eisert$^{1}$}

\affiliation{
$^1$Blackett Laboratory, Imperial College London, London SW7 2BW, UK\\
$^2$Institut~f{\"u}r~Laser-Physik,~Universit{\"a}t~Hamburg,~22761~Hamburg,~Germany}

\date\today

\begin{abstract}
Mixtures of bosonic and fermionic atoms in optical lattices provide a promising
arena to study strongly correlated systems. In experiments realizing such mixtures in the
quantum degenerate  regime the temperature is a key parameter. In this work, we investigate the 
intrinsic heating and cooling effects due to an entropy-preserving raising of the optical lattice potential. We analyze this process, identify the generic behavior valid for a wide range of parameters, and discuss it quantitatively for the recent experiments with $^{87}{\rm Rb}$ and $^{40}{\rm K}$ atoms.
In the absence of a lattice, we treat the bosons in the Hartree-Fock-Bogoliubov-Popov-approximation, including the fermions in a self-consistent mean field interaction. In the presence of the full three-dimensional lattice, we use a strong coupling expansion. 
As a result of the presence of the fermions, the temperature of the mixture after the lattice ramp-up is always higher than for the pure bosonic case. This sheds light onto a key point in the analysis of recent experiments.
\end{abstract}


\maketitle

Interacting bosonic and fermionic systems play a key role in 
several contexts in physics, quite prominently in the 
BCS theory of superconductivity. Systems of dilute atomic gases
(in optical lattices) offer the perspective of simulating such
mixtures or purely bosonic or fermionic systems under extraordinarily
controlled conditions \cite{reviews,fermions}.
Bose-Fermi mixtures in optical
lattices exhibit a rich physical behavior, including a wealth
of novel phases, charge density waves and 
supersolids \cite{BFHubbard}. Recent experiments have succeeded
in preparing such mixtures in optical lattices \cite{Inguscio}, notably the
realization of a stable bosonic $^{87}{\rm Rb}$ and
fermionic $^{40}{\rm K}$ mixture in three-dimensional 
optical lattices  \cite{Esslinger,OspelkausTheExp}.

To achieve realizations of such strongly correlated systems in the quantum-degenerate regime, very low temperatures have to be reached. This is not only a difficult prescription but also, while thermometry methods in the absence of a lattice are established, it is not entirely clear how to measure the temperature in its presence. Indeed, following recent experiments with cold bosonic atoms, an intriguing and fruitful controversy \cite{controversy} has arisen concerning the general question relevant to experiments with ultracold atoms in optical lattices: How cold, after all, is the system in the optical lattice expected to be? For Bose-Fermi mixtures, this question is even harder to answer as the additional degrees of freedom leave more room for different explanations. Interactions between bosons and fermions result in an effectively reduced repulsion between bosons, independent of the sign of the Bose-Fermi interaction. Hence, one might well expect an increase in coherence as compared to the purely bosonic case.
Quite surprisingly, however, the opposite effect (as measured in terms of the visibility of the quasi-momentum distribution) was observed \cite{Esslinger,OspelkausTheExp}. The theoretical work Ref.\ \cite{Kollath}, 
based on numerical quantum Monte Carlo and 
density-matrix-renormalization-group simulations
of one-dimensional systems, points towards the possibility that this might actually be due to a finite temperature effect.

In this work, we discuss the thermodynamics of adiabatic loading of  harmonically trapped Bose-Fermi mixtures into optical lattices. During the adiabatic loading procedure, the entropy remains constant and leads to intrinsic cooling or heating processes. We argue that one should expect a significant adiabatic heating of the mixture, not to be confused with experimental imperfections such as parametric heating. This is by no means a marginal effect, as we will quantitatively
clarify. This resulting temperature determines the physics of the strongly correlated system once the optical lattice is present. 

We identify the generic behavior and discuss it on the basis of the values corresponding to the experiment described in Ref.\ \cite{OspelkausTheExp}. More precisely, the presence of fermions leads either to a more distinct heating of the mixture or a less distinct cooling. We study in detail the behavior of these adiabatic heating and cooling effects in the 
inhomogeneous system---complementing results for purely {\hyphenchar\font=-1 (non-)}interacting bosonic  (\cite{BlakieBose}) \cite{other,Ho} and non-interacting fermionic \cite{BlakieFermi} systems---and analyse and flesh out the specific role of the fermions in this adiabatic process.

{\it Trapped Bose-Fermi mixture without optical lattice. --}
Subsequently, we will discuss the thermodynamics of the Bose-Fermi mixture in an isotropic harmonic trap in the absence of an optical lattice. We will insist on being close to an experimental situation in our description, and take the full three-dimensional situation  into account. We start from the grand-canonical Hamiltonian
\begin{equation}
\nonumber
\begin{split}
\hat{H}=&\int\!\md\vec{r}\,\hat{\Phi}^\dagger\hat{h}_B\hat{\Phi}+\int\!\md\vec{r}\,\hat{\Psi}^\dagger\hat{h}_F\hat{\Psi}
\\
&+\frac{g_{BB}}{2}\int\!\md\vec{r}\,
\hat{\Phi}^\dagger\hat{\Phi}^\dagger
\hat{\Phi}\hat{\Phi}
+g_{FB}\int\!\md\vec{r}\,
\hat{\Phi}^\dagger\hat{\Phi}
\hat{\Psi}^\dagger\hat{\Psi},
\end{split}
\end{equation}
where we denoted the bosonic (fermionic) field operators by $\hat{\Phi}$ ($\hat{\Psi}$), the interaction amplitudes $g_{BB}$, $g_{FB}$ are related to the respective scattering lengths as $g_{BB}=4\pi\hbar^2a_{BB}/m_B$, $g_{FB}=2\pi\hbar^2a_{FB}(m_B+m_F)/(m_Fm_B)$, 
the free part of the bosonic Hamiltonian is given by
\begin{equation}
\nonumber
\begin{split}
\hat{h}_{B}&=-\frac{\hbar^2}{2m_{B}}\vec{\nabla}^2
+V_{B}-\mu_{B},\;\;\;
V_{B}=\frac{m_{B}\omega_{B}^2}{2}\vec{r}^2,
\end{split}
\end{equation}
and accordingly for $\hat{h}_F$. We thus restrict ourselves to isotropic traps, taking geometrical averages of the trapping frequencies in the actual experiment \cite{OspelkausTheExp}.
 
\begin{figure}
\begin{center}
\includegraphics[width=\columnwidth]{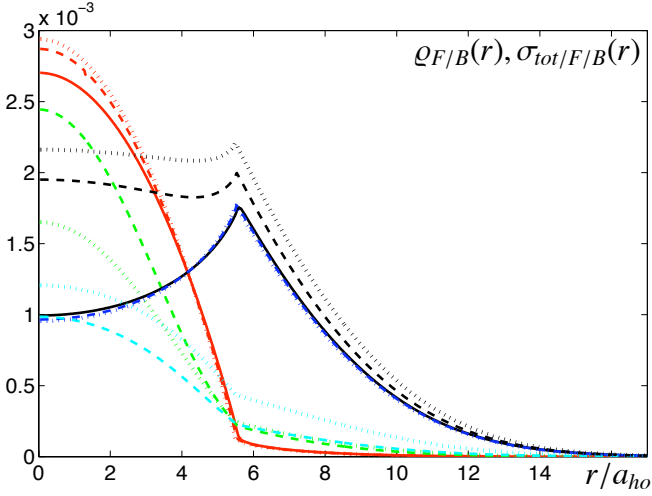}
\end{center}
\caption{Densities $\varrho_{B}=na_{ho}^3/N_B$ (red), $\varrho_{F}=ma_{ho}^3/N_F$ (green) and entropy densities $\sigma_{B/F}=s_{B/F}a_{ho}^3/10^{4}$ (blue/cyan, scaled by $10^{4}$ for clarity), $\sigma_{tot}=\sigma_B+\sigma_F$ (black),
in units of the bosonic harmonic oscillator length $a_{ho}$. 
The shown data sets correspond to the experiment in Ref.\ \cite{OspelkausTheExp} and a temperature of $95\,$nK. The mixture consists of $10^5$  $^{87}$Rb atoms and no (solid), $N_F=0.03N_B$ (dashed), and $N_F=0.07N_B$ (dotted) $^{40}$K atoms.}\label{Fig1}
\end{figure}
For the bosonic sector, we invoke the standard 
local density Hartree-Fock-Bogoliubov-Popov (HFBP) approximation, 
which is a
self-consistent mean field scheme that has proven applicable to a wide
temperature regime, see Refs.\ \cite{Stringari,Stringari2} and references therein. The inter-species interaction is treated in
the 
self-consistent mean-field approximation (see, e.g., Refs.\ \cite{mixture_thermo,Pelster})
\begin{equation}
\nonumber
\begin{split}
\hat{\Phi}^\dagger\hat{\Phi}
\hat{\Psi}^\dagger\hat{\Psi}&\approx
\hat{\Phi}^\dagger\hat{\Phi}
\langle\hat{\Psi}^\dagger\hat{\Psi}\rangle+\langle\hat{\Phi}^\dagger\hat{\Phi}\rangle
\hat{\Psi}^\dagger\hat{\Psi}-\langle\hat{\Phi}^\dagger\hat{\Phi}\rangle\langle
\hat{\Psi}^\dagger\hat{\Psi}\rangle\\
&=:\hat{\Phi}^\dagger\hat{\Phi}
m+n
\hat{\Psi}^\dagger\hat{\Psi}-mn,
\end{split}
\end{equation}  
where we defined the fermionic density $m$ and the total bosonic density $n=n_0+n_T$ composed of the condensate and non-condensate density $n_0$, $n_T$, respectively. This yields the following
set of coupled equations: (i) The finite temperature Gross-Pitaevskii
equation in the Thomas-Fermi approximation (which may be safely applied for the high number of atoms considered), governing the condensate density
\begin{equation}
\nonumber
n_0=\max\left\{0,\frac{\mu_B-V_B-g_{FB}m}{g_{BB}}-2n_T\right\},
\end{equation}
where the chemical potential is fixed by the given total number of bosons, $N_B=N_0+N_T=\int\!\md\vec{r}\,n_0+\int\!\md\vec{r}\,n_T$. (ii) The thermal density of bosons ($k_BT=1/\beta$)
\begin{equation}
\nonumber
n_T=\int\!\frac{\md\vec{p}}{(2\pi)^3}\left[\frac{u^2_+(\vec{p},\vec{r})+u^2_-(\vec{p},\vec{r})}{\me^{\beta\epsilon(\vec{p},\vec{r})}-1}+u^2_-(\vec{p},\vec{r})\right],
\end{equation}
where the Bogoliubov amplitudes are given by
\begin{equation}
\nonumber
u_{\pm}^2=\frac{\frac{\hbar^2\vec{p}^2}{2m_B}+V_B-\mu_B+2g_{BB}n+g_{FB}m}{2\epsilon}\pm\frac{1}{2},
\end{equation}
and the quasi-particle spectrum reads
\begin{equation}
\nonumber
	\epsilon^2={\left(\frac{\hbar^2\vec{p}^2}{2m_B}+V_B-\mu_B
	+2g_{BB}n+g_{FB}m\right)^2-g_{BB}^2n_0^2}.
\end{equation}
Finally, (iii) the fermionic density in local density approximation
\begin{equation}
\nonumber
m=\int\!\frac{(2\pi)^{-3}\md\vec{p}}{\me^{\beta\delta(\vec{p},\vec{r})}+1},\;\;\;
\delta=\frac{\hbar^2\vec{p}^2}{2m_F}+V_F-\mu_F+g_{FB}n,
\end{equation}
where the chemical potential is fixed by the given total number of fermions $N_F=\int\!\md\vec{r}\,m$. 

For given temperature $T$ and particle numbers $N_B$, $N_F$, we solve (i)-(iii) self-consistently 
in the following way: Starting with no interaction between 
bosons and fermions, $g_{FB}=0$ and $n_T=0$, 
we (a) compute $n_0$ and $\mu_B$ by solving (i) under the particle number restriction, (b) obtain $n_T$ from (ii), (c) iterate (a) and (b) until convergence, (d) solve (iii), which yields $m$ and $\mu_F$, (f) iterate (a)-(d) until convergence.

After convergence, we are equipped with the energies $\epsilon$, $\delta$, and can compute the entropy of the mixture,
\begin{equation}
\nonumber
S/k_B=\int\!\frac{\md\vec{p}\md\vec{r}}{(2\pi)^3}\left[s_B(\vec{p},\vec{r})+s_F(\vec{p},\vec{r})\right],
\end{equation}
where the individual contributions read
\begin{equation}
\nonumber
\begin{split}
s_B(\vec{p},\vec{r})&=\frac{\beta\epsilon(\vec{p},\vec{r})}{\me^{\beta\epsilon(\vec{p},\vec{r})}-1}-\log\left(1-\me^{-\beta\epsilon(\vec{p},\vec{r})}\right),\\
s_F(\vec{p},\vec{r})&=\frac{\beta\delta(\vec{p},\vec{r})}{\me^{\beta\delta(\vec{p},\vec{r})}+1}+\log\left(1+\me^{-\beta\delta(\vec{p},\vec{r})}\right).
\end{split}
\end{equation}
This is the expression forming the starting point of the analysis
in the absence of the lattice. In Fig.\ \ref{Fig1}, we show the obtained
results for the parameters of the experiments in Ref.\ \cite{OspelkausTheExp} for  different ratios $N_F/N_B$. The critical temperature for Bose condensation is $\approx 205\,$nK for all three cases $N_F/N_B=0, 0.03, 0.07$. In this experiment no thermal cloud was discernible,
corresponding to a BEC fraction of at least $80$\% and a initial temperature below $95\,$nK. The bosonic entropy is highest at the condensate boundary, where the density of the thermal cloud has its maximum. In turn, $s_F$ is highest in the center of the trap. We can see that the bosonic contribution, $s_B$, to the total entropy remains basically unaltered by the presence of the fermions, their main contribution to $S$ stemming from $s_F$ itself.

{\it Trapped mixture in deep optical lattices. --}
To describe the system in the presence of the lattice, we use the  single-band Bose-Fermi-Hubbard Hamiltonian 
\cite{BFHubbard,amplitudes} $\hat{H}=\hat{J}+\sum_i\hat{h}_i$,
where
\begin{equation}
\nonumber
\begin{split}
\hat{J}&=-J_F\sum_{\langle i,j\rangle}\hat{f}_i^\dagger\hat{f}_j-J_B\sum_{\langle i,j\rangle}\hat{b}_i^\dagger\hat{b}_j,\\
\hat{h}_i&=\frac{U}{2}\hat{n}_i(\hat{n}_i-1)+V\hat{n}_i\hat{m}_i-\mu^B_i\hat{n}_i-\mu^F_i\hat{m}_i.
\end{split}
\end{equation}
Here, the operator $\hat{b}_i$ ($\hat{f}_i$) annihilates a boson (fermion) at site $i$ and $\hat{n}_i=\hat{b}_i^\dagger\hat{b}_i$,  $\hat{m}_i=\hat{f}_i^\dagger\hat{f}_i$. $\hat{J}$ accounts for the tunneling of atoms from one site to neighboring sites, $U$, $V$ are the intra-, respectively inter-species on-site interactions, and $\mu_i^{B/F}=\mu_{B/F}-V_i$ are on-site chemical potentials controlling the particle number via $\mu_{B/F}$ and accounting for the harmonic confinement $V_i$, which is approximately the same for both species. For deep lattices the tunneling may be treated as a perturbation in $J_{B/F}$. Up to second order and within local density approximation (assuming that the 
trapping potentials are the same at neighboring sites), the free energy is found to be $F=-\log(Z)/\beta=-\sum_i\log(z_i)/\beta-3\sum_i(J_B^2b_i+J_F^2f_i)/z_i^2$, where 
$b_i=\sum_{n,n'=0}^\infty n(n'+1)b^{n,n'}_i$,
\begin{eqnarray}
b^{n,n'}_i\!\!&=&\!\!\!\!\sum_{m,m'=0}^1\!\!\!\!\me^{-\beta(\epsilon^{n,m}_i+\epsilon^{n',m'}_i)}\frac{\me^{\beta[U(n'-n-1)+V(m'-m)]}-1}{U(n'-n-1)+V(m'-m)},\nonumber\\
f_i&=&\!\!\!\!\sum_{n,n'=0}^\infty\!\!\!\!\me^{-\beta(\epsilon^{n,1}_i+\epsilon^{n',0}_i)}
\frac{\me^{\beta V(n'-n)}-1}{V(n'-n)}\nonumber,
\end{eqnarray}
and the unperturbed on-site energies and corresponding partition functions are given by 
$\epsilon^{n,m}_i=Un(n-1)-\mu_i^Bn-\mu_i^Fm+Vnm$,
$z_i=\sum_{n=0}^\infty[\exp(-\beta\epsilon^{n,0}_i)+\exp(-\beta\epsilon^{n,1}_i)]$. The structure of a sum over on-site terms originates from the local-density approximation and helps to speed up the computation substantially: All sites with the same value $V_i$ yield the same contribution to the free energy. They may thus be grouped together and the contribution be calculated only once per group as opposed to for every single site. 
Starting from the above  expression for the free energy, 
we calculate the chemical potentials for given particle numbers $N_{B/F}$ by numerically solving 
$N_{B/F}=-\partial F/\partial\mu_{B/F}$, where the right hand side is obtained by numerically differentiating the free energy with respect to the chemical potentials.  Similarly, by differentiating with respect to $\beta$, we then compute the entropy of the mixture in the lattice: 
$S/k_B=\beta^2\partial F/\partial\beta$.

\begin{figure}
\begin{center}
\includegraphics[width=\columnwidth]{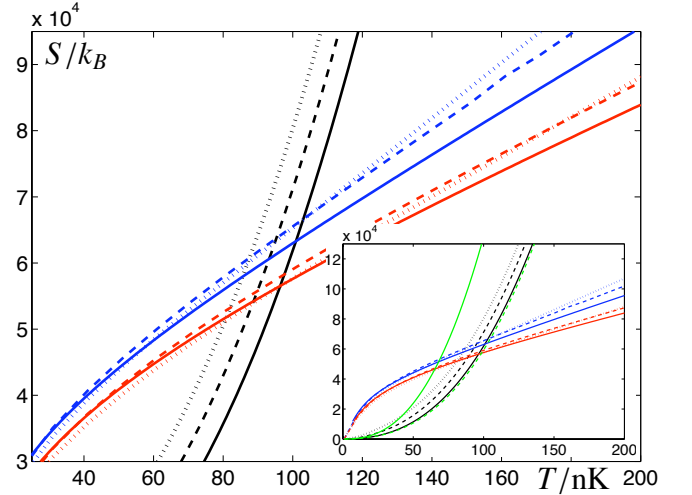}
\end{center}
\caption{Entropy as a function of temperature for the parameters of Ref.\ \cite{OspelkausTheExp}. The mixture consists of $10^5$  $^{87}$Rb atoms and $N_F=0$ (solid), $0.03N_B$ (dashed), $N_F=0.07N_B$ (dotted) $^{40}$K atoms in a lattice of various depths (blue: $15 E_R$, red: $30 E_R$, black: no lattice).
The inset shows the same at a larger scale, including the entropy for the purely bosonic case obtained from analytical expressions (green) valid below the critical temperature (dashed) and
at ultra-low $T$ (solid), see footnote \cite{Ho_comment}.}\label{Fig2}
\end{figure}

\begin{figure}
\begin{center}
\includegraphics[width=\columnwidth]{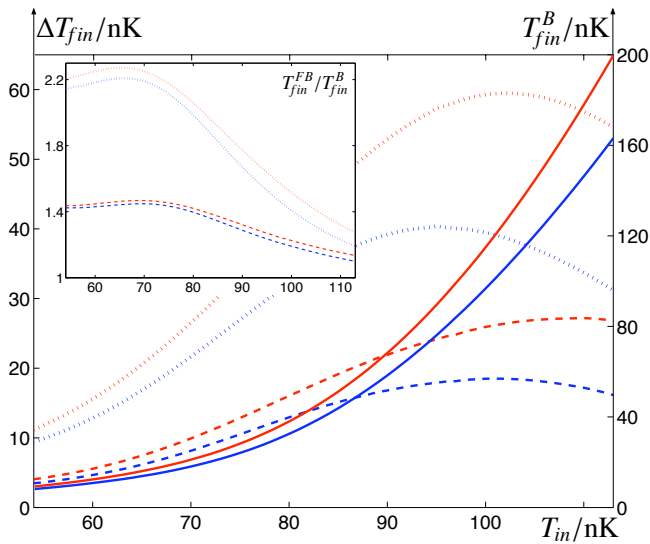}
\end{center}
\caption{Difference $\Delta T_{\!fin}=T_{\!\!fin}^{FB}-T_{\!fin}^B$ between the final temperature in the lattice with and without fermions as a function of the initial temperature without lattice. Parameters are as in Fig.\ \ref{Fig2}. For any initial temperature, the presence of the fermions leads to higher final temperatures as compared to the purely bosonic case. Solid lines depict the final temperature $T_{\!fin}^B$ without fermions (right scale).}\label{Fig3}
\end{figure}

{\it Discussion. --} We are now in the position to assess the situation when ramping up the lattice in an adiabatic process. Fig.\ \ref{Fig2} shows the entropy as a function of temperature in a
system of $10^5$ $^{87}$Rb-atoms in a three-dimensional trap. In this figure, we
use the experimental parameters of Ref.\ \cite{OspelkausTheExp}, but the findings are valid
for a wide range of parameters. Both the entropy without and in the presence of the optical lattice is 
depicted for the purely bosonic case and a small admixture of fermions $N_F/N_B=0.03, 0.07$. 
We see that generally, for fixed lattice depth, below a certain temperature, the adiabatic ramp-up gives rise to an adiabatic cooling, whereas above this threshold temperature we find an adiabatic heating. Both above and below this temperature, the presence of the fermions results in a higher final temperature as compared to the same situations with bosons only. This is most dramatic at initial temperatures for which without fermions the bosons are adiabatically cooled and in the mixture adiabatic heating occurs, e.g., for an initial temperature of $90\,$nK and a final lattice depth of $15E_R$ the temperature is $\approx 40\,$nK higher in the presence of $N_F=0.07N_B$ fermions, corresponding to an increase of $\approx 67\,$\%, see Fig.\ \ref{Fig3}. This affects the contrast of the interference pattern \cite{Kollath} analyzed in those experiments. This behavior is  generic, valid in particular for both experiments 
of Refs.\ \cite{Esslinger,OspelkausTheExp} as well as for experiments performed in an isotropic and shallower trap \cite{OspelkausThesis}, where the initial temperature was always below the threshold: For any initial temperature, the entropy without the lattice is always much higher in the presence of fermions, even for the relatively small admixture of $^{40}$K atoms as in Fig.\ \ref{Fig2}. While below the threshold adiabatic cooling occurs, this effect is lessened compared to the purely bosonic case. Above the threshold and in the  presence of the lattice, the entropy including fermions is higher, thus reducing the heating effect. This can however not compensate for the high initial difference of entropies, see Fig.\ \ref{Fig3}. 

Note that the influence of fermions is most distinguished in the absence of the lattice. This is plausible when considering the form of the unperturbed free energy in the presence of the lattice: $\epsilon_i^{n,0}= U n(n-1) - \mu_i^B n$ and $\epsilon_i^{n,1}= U n(n-1) - \mu_i^B n + V n$
are different only by an alteration of a definition of the bosonic chemicals potentials (the total number of bosons is the same with and without fermions), leading for low temperatures to approximately the same expression for the entropy. Taking a closer look at the situation including the lattice, we see that, at low temperatures, more fermions lead to a lower entropy -- the inter-species
attraction reducing the mobility of the atoms and thus reducing the
number of possible micro-states. In turn, at higher temperatures interactions become less important and the entropy increases with the number of fermions. While this can not compensate the initial difference in entropies, it however reduces the heating effect for higher initial temperatures, see Fig.\ \ref{Fig3}. This effect may be observed in the currently available experiments: At a fixed lattice depth, the difference between the situation with and without fermions should first increase, reach a maximum and finally decrease with increasing initial temperature.

{\it Summary and outlook. --} In this work, we have quantitatively explored the adiabatic cooling and heating effects that are to be expected in experiments with Bose-Fermi mixtures in optical lattices, crucial when reaching a strongly correlated system. On intuitive grounds, one could have suspected that the features observed in experiments were entirely due to a shift of the bosonic Mott lobes in the presence of fermions, the presence of fermions effectively altering the local chemical potential. This is indeed the case, but predicts an increase of coherence \cite{Kollath}, the opposite of which was observed in experiments. We have seen that under the parameters as used in present experiments, the resulting temperature is much larger than expected from thermometry based on measurements before the ramp-up of the lattice. 
Methods to assess the temperature of samples within deep optical lattices would clearly be a breakthrough for any studies on quantum gases in lattices. Promising ideas are, e.g., the detailed characterization of the shell structure of local densities \cite{Foelling}.
A link to the expected visibility from our analysis is provided by Ref.\ \cite{Kollath}. 
This analysis applies to a one-dimensional situation, yet for the visiblility, it is expected to give a 
clear guideline: It is seen how the bosonic visibility decreases as the temperature 
increases. A clear-cut quantitative analytical analysis of the quasi-momentum distribution at finite temperature is still lacking and poses---even for purely bosonic systems---an exciting challenge and constitutes a test-bed for theories developed in the condensed matter context. It is the hope that the present work can significantly contribute to the clarification of the intriguing discussion on the interpretation of observed data and on the available theoretical models for ultracold mixtures of bosonic and fermionic atoms in optical lattices.

{\it Acknowledgements. --} This work has been supported by the DFG (SPP 1116), the EU (QAP), Microsoft Research, EURYI, and the EPSRC. We would like to thank A.\ Chudnovskiy, C.\ Kollath, and
I.\ Bloch for fruitful discussions.


\begin{thebibliography}{99}


\bibitem{reviews}
	I.\ Bloch, J.\ Dalibard, and W.\ Zwerger,
	arXiv:0704.3011v1;
	D.\ Jaksch and P.\ Zoller, 
	cond-mat/0410614v1;
	M.\ Lewenstein et al., 	
	Adv.\ Phys.\ {\bf 56}, 243 (2007).
 	
\bibitem{fermions}
	C.A.\ Regal, M.\ Greiner, and D.S.\ Jin,
	Phys.\ Rev.\ Lett.\ {\bf 92}, 040403 (2004);
	M.W.\ Zwierlein et al.,
	ibid.\ {\bf 92}, 120403 (2004);
	T.\ Bourdel et al.,
	ibid.\ {\bf 93}, 050401 (2004);
	J.\ Kinast et al.,
	Science  {\bf 307}, 1296 (2005);
	C.\ Chin et al., 
	ibid.\ {\bf 305}, 1128 (2004);
	M.\ K{\"o}hl et al,
	Phys.\ Rev.\ Lett.\ {\bf 94}, 080403 (2005);
	T.\ Rom et al., Nature {\bf 444}, 733 (2006).

\bibitem{BFHubbard}
	A.\ Albus, F.\ Illuminati, and J.\ Eisert,
	Phys.\ Rev.\ A {\bf 68},  023606 (2003);
	M.\ Lewenstein et al., Phys.\ Rev.\ Lett.\ {\bf 92}, 050401 
	(2004); 	
	M.\ Cramer, J.\ Eisert, and F.\ Illuminati, 
	ibid.\ {\bf 93}, 190405 (2004);
	L.\ Mathey et al., 
	ibid.\ {\bf 93}, 120404 (2004);
	F.\ Illuminati and A.\ Albus, ibid.\ {\bf 93}, 090406 (2004);
	H.P.\ B{\"u}chler and G.\ Blatter, 
	Phys.\ Rev.\ A {\bf 69}, 063603 (2004);
	R.\ Roth and K.\ Burnett, 
	ibid.\ {\bf 69}, 021601(R) (2004);	
	E.\ Pazy and A.\ Vardi, ibid.\ {\bf 72}, 033609 (2005).	
	
	\bibitem{Inguscio}
	F.\ Ferlaino et al., 
	Phys.\ Rev.\ Lett.\ {\bf 92}, 140405 (2004).
	
	\bibitem{Esslinger}
	K.\ G{\"u}nter et al.,
	Phys.\ Rev.\ Lett.\ {\bf 96}, 180402 (2006).
	
\bibitem{OspelkausTheExp}
	 S.\ Ospelkaus et al.,	
	 Phys.\ Rev.\ Lett.\ {\bf 96}, 180403 (2006).
	 The experiment was performed for a mixture of $10^5$ $^{87}$Rb ($a_{BB}=5.238\,$nm) and $^{40}$K ($a_{FB}=-10.848\,$nm) atoms confined in a magnetic trap with $\omega_x=\omega_y=2\pi\cdot 150\,$Hz, and $\omega_z=2\pi\cdot 50\,$Hz, subject to a three-dimensional lattice formed by lasers with a wavelength of $1030\,$nm and waists $w_x=82\,\mu$m, $w_y=92\,\mu$m, and $w_z=55\,\mu$m, which results in an additional harmonic confinement according to $\omega_i^2\rightarrow\omega_i^2+8V_0/(m_Bw_i^2)$.
	 		
\bibitem{controversy}
	R.\ B.\ Diener et al., Phys.\ Rev.\ Lett.\ {\bf 98}, 180404 	(2007); 
	F.\ Gerbier et al., cond-mat/0701420.
			
\bibitem{Kollath}
	L.\ Pollet et al., cond-mat/0609604.	
	
\bibitem{BlakieBose}	
	 P. B.\ Blakie and J.V.\ Porto, 
	 Phys.\ Rev.\ A {\bf 69}, 013603 (2004).
	 
\bibitem{other}
	B.\ Capogrosso-Sansone, N.V.\ Prokof'ev, and B.V. Svistunov,
	 Phys.\ Rev.\ B	{\bf 75}, 134302 (2007).
	

\bibitem{Ho}
	T.-L.\ Ho and Q.\ Zhou, cond-mat/0703169.

\bibitem{BlakieFermi}
 	P.B.\ Blakie and A.\ Bezett, Phys.\ Rev.\ A 
	{\bf 71}, 033616 (2005).
	
\bibitem{OspelkausThesis}
	S. Ospelkaus, PhD thesis, 
	University of Hamburg (2006).
	
\bibitem{Ho_comment}
	The ultra-low temperature approximation for the energy 
	$E/(N_Bk_BT_c^0)=5\eta/7+10.6\eta^{1/2}t^{7/2}$ 
	\cite{Stringari2} 
	used in Ref.\ \cite{Ho} to obtain the entropy is somewhat 
	misleading in its predictions. Comparing it to the local-density 
	HFBP result shows that for 
	the considered parameter and temperature 
	regime the expression	
	$E/(N_Bk_BT_c^0)=3\zeta(4)t^4/\zeta(3)+\eta (1-t^3)^{2/5}(5+16t^3)/7$
	 \cite{Stringari} is more appropriate, see also 
	 Ref.\ \cite{other}. Here, 
	$\eta=\zeta(3)^{1/3}15^{2/5}(N_B^{1/6}a_{BB}/a_{ho})^{2/5}/2$, 
	$t=T/T_c^0$, and 
	$T_c^0$ 
	is the 
	critical temperature in absence of interactions.
	In Fig.\ \ref{Fig2} we depict the entropy obtained from the 
	above expression (green dashed line) and the one used in 
	Ref.\ \cite{Ho} (green solid line). 
	For the parameters in Ref.\ \cite{Ho} we have 
	also compared both expressions 
	to the local-density HFBP result (both former expressions are 
	approximations to the latter) and 
	found the same fundamental difference. Interestingly, 
	this refined study
	shows that instead of the claimed adiabatic heating, 
	cooling should occur for typical initial temperatures. 
	This 
	is also substantiated by the observation of the spatial 
	shell structures in Ref.\ \cite{Foelling}, which would not be 
	visible in the case of
	heating.

\bibitem{Foelling}
	S. Foelling et al., 
	Phys.\ Rev.\ Lett.\ {\bf 97}, 060403 (2006).
	
\bibitem{Stringari}
	F.\ Dalfovo et al., Rev.\ Mod.\ Phys.\
	{\bf 71}, 463 (1999).
	
\bibitem{Stringari2}
	S.\ Giorgini, L.P.\ Pitaevskii, and S.\ Stringari,
	J.\ Low Temp.\ Phys.\ {\bf 109}, 309 (1997).
	
\bibitem{Burnett}
	R.J.\ Dodd et al., Phys.\ Rev.\ A {\bf 57},  
	R32 (1998).
	
\bibitem{mixture_thermo} 
	H.\ Hu and X.-J.\ Liu, Phys.\ Rev.\ A {\bf 68},
	023608 (2003). 

\bibitem{Pelster}
	S.\ R{\"o}thel and A.\ Pelster, 
	cond-mat/0703220.
	
\bibitem{amplitudes}
	We obtain the amplitudes by numerically computing 
	the exact single particle Wannier functions and 
	their overlap.
		
	
\end{thebibliography}
\end{document}